\documentclass[twocolumn,aps,prl,amsmath,floatfix]{revtex4}
\usepackage{graphicx}
\begin{document}
\title{The Mott insulator LaTiO$_3$ in heterostructures with SrTiO$_3$ is metallic} 
\author{H. Ishida$^1$ and A. Liebsch$^2$}
\affiliation{
\mbox{$^1$College of Humanities and Sciences, Nihon University, and CREST JST,
        ~Tokyo 156, Japan}          \\
\mbox{$^2$Institut f\"ur Festk\"orperforschung,~Forschungszentrum J\"ulich,
        ~52425 J\"ulich, Germany} 
}
\begin{abstract}
It is shown that LaTiO$_3$ in superlattices with SrTiO$_3$ is not a 
Mott insulator but a strongly correlated metal. The tetragonal lattice 
geometry imposed by the SrTiO$_3$ substrate leads to an increase of the 
Ti $3d$ $t_{2g}$ band width and a reversal of the $t_{2g}$ crystal field 
relative to the orthorhombic bulk geometry. Using dynamical mean field theory 
based on finite-temperature multi-band exact diagonalization we show that, 
as a result of these effects, local Coulomb interactions are not strong 
enough to induce a Mott transition in tetragonal LaTiO$_3$. The metalicity 
of these heterostructures is therefore not an interface property but 
stems from all LaTiO$_3$ planes.
\\ \mbox{\ }\\     
DOI: \hfill PACS numbers: 73.21.-b, 71.27.+a., 73.40.-c 78.20.-e
\end{abstract}
\maketitle

There exists currently considerable interest in the design of 
nano-materials with electronic properties that differ qualitatively from
those of the constituent components in their bulk form. A particularly
intriguing example is the formation of a thin metallic layer at the  
interface between two insulators. For instance, SrTiO$_3$ is a band 
insulator with an empty $d$ band, whereas LaTiO$_3$, with one $d$ electron
per site, is regarded as a textbook Mott insulator because of strong local 
Coulomb interactions~\cite{imada}.  
Nevertheless, the pioneering work by Ohtomo {\rm et al.}~\cite{ohtomo} 
shows that LaTiO$_3$/SrTiO$_3$ superlattices are metallic, where the overall 
conductivity depends on the thickness of the LaTiO$_3$ interlayers and on 
the spacing between them. This phenomenon 
appears to follow from the fact that the Ti interface layer between adjacent 
La and Sr planes formally exhibits a $3d^{0.5}$ valency, in contrast to the 
$3d^1$ and $3d^0$ configurations of bulk LaTiO$_3$ and SrTiO$_3$, respectively.
Other examples are LaAlO$_3$/SrTiO$_3$ heterostructures where a conducting 
interface was observed although both constituents are wide band gap 
perovskite insulators~\cite{hwang}. 
Very recently, various other perovskite superlattices have been investigated
experimentally~\cite{takizawa,kourkoutis,maekawa,wadati,may,bhattacharya,hotta,sheets}.

One of the hallmarks of transition metal oxides is their extreme sensitivity
to small changes of key parameters such as temperature, pressure, impurity 
concentration, or structural distortion~\cite{imada}. 
The aim of this work is to demonstrate that, as a result of a subtle change 
of crystal structure, LaTiO$_3$ in superlattices with SrTiO$_3$ is a strongly 
correlated metal, rather than a Mott insulator like bulk LaTiO$_3$. 
As a consequence, the metalicity observed in~\cite{ohtomo} is not only a 
property of the interface but of the entire LaTiO$_3$ layer. 

The origin of this symmetry induced insulator to metal transition is that, 
when LaTiO$_3$ is grown on cubic SrTiO$_3$, the substrate imposes a tetragonal 
geometry on the first few layers~\cite{kim}. 
As we show below, this structural modification leads to two significant changes
of the electronic properties. First, it causes a substantial increase of the 
Ti $t_{2g}$ band width. Second, the $t_{2g}$ crystal field is weaker and has 
the opposite 
sign compared to bulk LaTiO$_3$. Both effects ensure that the Mott transition 
in tetragonal LaTiO$_3$ occurs at a significantly larger critical Coulomb 
energy $U_c$ than in bulk LaTiO$_3$. Thus, for realistic values of $U$, 
thin LaTiO$_3$ layers in heterostructures with SrTiO$_3$ are strongly 
correlated metals.
These results underline the importance of carefully characterizing the 
interfaces of superlattices that might be employed in future devices.  

The electronic properties of LaTiO$_3$/SrTiO$_3$ heterostructures have been 
studied theoretically within a variety of single-electron and many-electron 
models~\cite{okamoto.04,freericks,popovic.05,okamoto.06,hamann.06,lee.06,%
kancharla.07,pentcheva.07,ruegg.07}.
The conditions under which LaTiO$_3$ in these superlattices might be a Mott 
insulator, however, have not yet been explored. In view of the striking 
interplay between Ti $3d$ $t_{2g}$ orbital degrees of freedom and local 
Coulomb interactions this issue is crucial and might differ considerably
from the situation in bulk LaTiO$_3$. As shown by 
Pavarini {\it et al.}~\cite{pavarini}, the noncubic octahedral distortions in 
orthorhombic bulk LaTiO$_3$ give rise to nondiagonal components of the $t_{2g}$ 
density of states. Nevertheless, from these orbitals a new basis
can be constructed (denoted here as $a_g,\,e'_g$), in which the local density
of states is nearly diagonal, and where the $a_g$ contribution lies about 
200~meV below the nearly degenerate pair of $e'_g$ components, in qualitative
agreement with experimental findings~\cite{cwik}.  The local Coulomb energy 
greatly enhances this Ti $3d$ orbital polarization so that for $U\approx5$~eV 
a Mott transition occurs where the $a_g$ band is nearly half-filled and the 
$e'_g$ states are nearly empty~\cite{pavarini}. The important conclusion 
from this picture is that the metal insulator transition
in LaTiO$_3$ is intimately connected to the sign and magnitude of the 
$t_{2g}$ crystal field splitting which reflects, in turn, the orientation 
and magnitude of the noncubic lattice distortion.

Here we combine {\it ab initio} electronic structure calculations for the 
SrTiO$_3$ induced tetragonal geometry of LaTiO$_3$ with finite-temperature 
dynamical mean field theory (DMFT)~\cite{dmft,DMFT} and compare the correlated 
electronic properties with those of the bulk orthorhombic structure. 
As impurity solver we use a recent multi-band 
extension~\cite{perroni} of exact diagonalization (ED)~\cite{ed} which 
has been shown to be highly useful for the study of strong 
correlations in several transition metal oxides~\cite{perroni,CRO,NCO,doping}. 
Since this 
approach does not suffer from sign problems the full Hund exchange including 
spin-flip and pair-exchange contributions can be taken into account. Also, 
relatively large Coulomb energies and low temperatures can be reached.        

\begin{figure}[t]
  \begin{center}
  \includegraphics[width=5.0cm,height=8cm,angle=-90]{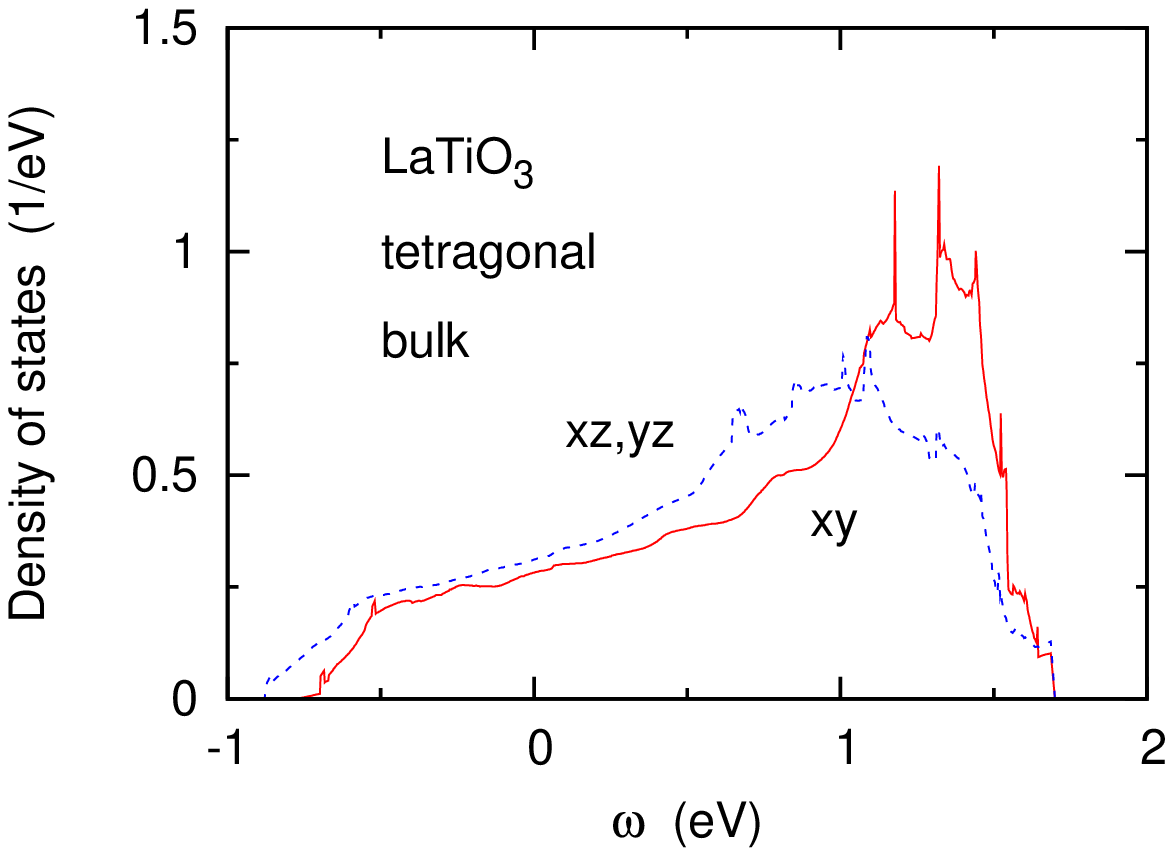}
  \includegraphics[width=5.0cm,height=8cm,angle=-90]{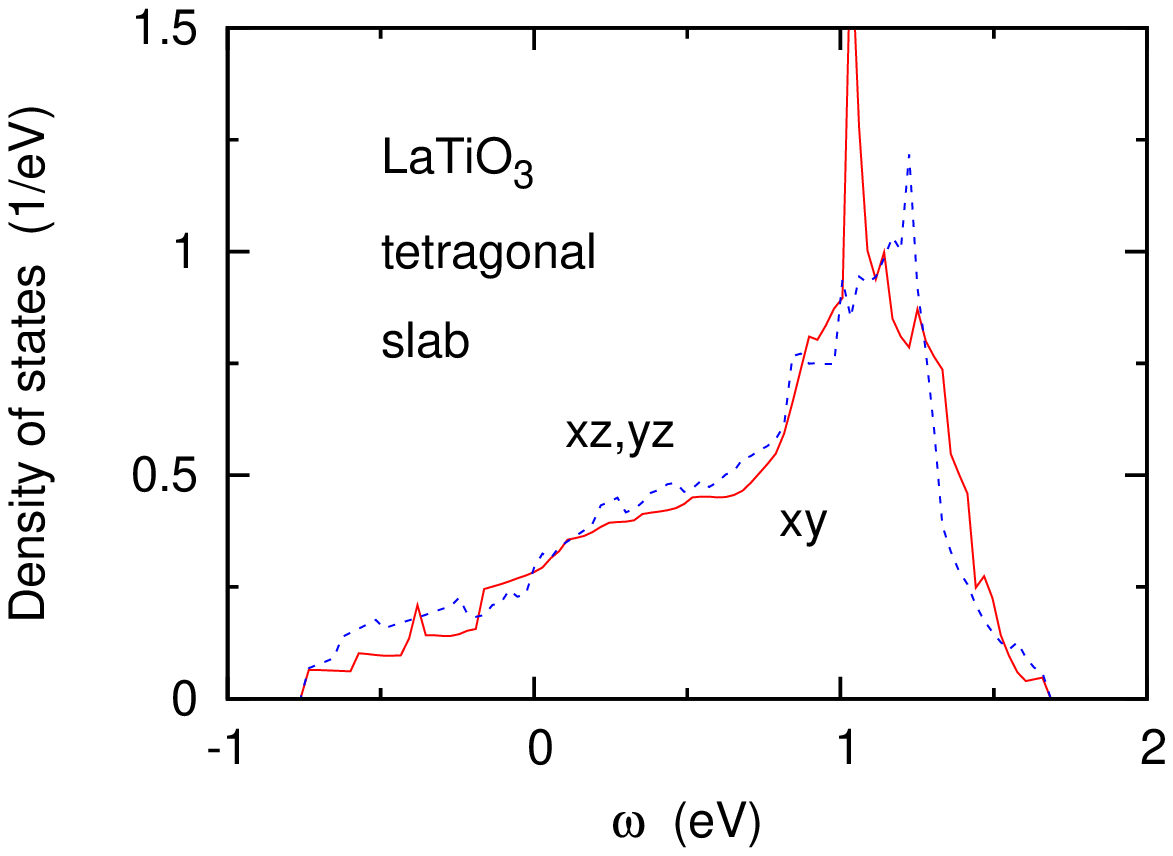}
  \end{center}
\vskip-4mm
\caption{(Color online) 
Upper panel: $t_{2g}$ density of states components LaTiO$_3$ in hypothetical 
tetragonal bulk geometry. Solid (red) curve: $d_{xy}$ states, dashed (blue) 
curves: $d_{xz,yz}$ states. $E_F=0$.
Lower panel: analogous density of states components for central Ti atom in 
(LaTiO$_3$)$_6$/(SrTiO$_3$)$_4$  superlattice (see text). 
}\end{figure}

The structures of LaTiO$_3$ in the tetragonal bulk and superlattice geometries
were optimized via the Vienna Ab-initio Simulation Package (VASP)~\cite{Kresse},
an implementation of the projector augmented-wave method~\cite{Blochl}, where 
the total energy of the system was calculated within non-spin-polarized GGA. 
The in-plane lattice constant was fixed to that of SrTiO$_3$, 
$a=b=3.92$~\AA, whereas
the lattice parameters in the $z$ direction were fully optimized. We assumed
a 1$\times$1 geometry without considering rotation and tilting of TiO$_6$
octahedra. For each optimized geometry, we performed an LDA calculation using
a home-made FLAPW code to calculate the partial density of states projected
on Ti $3d$ orbitals in muffin-tin spheres. For the bulk structure, the $z$ 
axis layer spacing $c=4.01$~\AA\ is only slightly smaller than $c=4.106$~\AA\ 
obtained for a tetragonal lattice with the same unit cell volume as in the 
orthorhombic case.

Fig.~1 shows the Ti $t_{2g}$ density of states components for the hypothetical
tetragonal bulk geometry of LaTiO$_3$ (upper panel). In this symmetry, the 
singly degenerate $d_{xy}$ density differs from the doubly degenerate 
$d_{xz,yz}$ densities. 
In the experimental work of Ohtomo {\it et al.}~\cite{ohtomo}, rather thin 
LaTiO$_3$ layers containing up to five La planes were studied. 
The lower panel of Fig.~1 shows the $t_{2g}$ density of states components 
for a (LaTiO$_3$)$_6$/(SrTiO$_3$)$_4$ 
superlattice, containing five Ti layers in a nominal $d^1$ configuration
(only La neighboring planes), three $d^0$ Ti layers (only Sr neighboring 
planes), and two $d^{0.5}$ Ti interface layers (La and Sr neighboring 
planes). Plotted is the density of states for the central Ti atom 
within the LaTiO$_3$ slab. The vertical height of the O$_6$ octahedron 
surrounding this atom, 3.98 \AA, is slightly smaller than that 
in the tetragonal bulk, 4.01 \AA. As a result, the difference between 
the $d_{xy}$ and $d_{xz,yz}$ densities is even smaller than that in 
bulk case shown in the upper panel of Fig. 1.

The overall width of the $t_{2g}$ bands for both tetragonal structures is 
about $W_{\rm tetra}\approx 2.5$~eV, where we have neglected small additional 
spectral weight in the region of the empty $e_g$ bands at higher energies.
Also, the centroid of the singly degenerate $d_{xy}$ band is seen to lie
above that of the $d_{xz,yz}$ bands. As a result of this {\it positive}
crystal field splitting, the subband occupancies (per spin band) are      
$n_{xy} \approx 0.14, \ n_{xz,yz} \approx 0.18 $ for the tetragonal bulk and  
$n_{xy} \approx 0.15, \ n_{xz,yz} \approx 0.175$ for the tetragonal slab. 

\begin{figure}[t]
  \begin{center}
  \includegraphics[width=5cm,height=8cm,angle=-90]{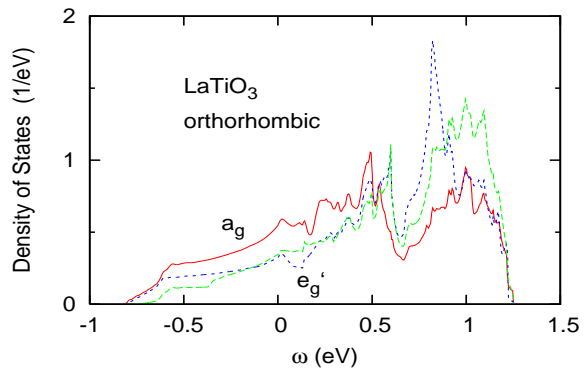}
  \end{center}
\vskip-4mm
\caption{(Color online) 
Density of states components of orthorhombic bulk LaTiO$_3$ in nearly 
diagonal subband representation~\cite{pavarini}. Solid (red) curve: 
$a_g$ states, dashed (blue and green) curves: $e'_g$ states. $E_F=0$.
}\end{figure}

For comparison we show in Fig.~2 the $a_g$ and $e'_g$ densities of orthorhombic 
bulk LaTiO$_3$~\cite{pavarini} which are obtained from a linear transformation 
of the original Ti $d_{xy,xz,yz}$ orbitals. 
As mentioned above, in the $a_g,\,e'_g$ basis the orthorhombic 
density of states is nearly diagonal and the centroid of the $a_g$ density
lies about 200~meV below the two $e'_g$ components, implying a {\it negative}
crystal field splitting between the singly and doubly degenerate densities.
The subband occupancies (per spin band) are $n_{a_g}\approx 0.23$ and 
$n_{e'_g}\approx 0.135$. Thus, orbital polarization is much larger than
for the tetragonal structures shown in Fig.~1 and has the opposite sign. 
Moreover, the width of the $a_g,\,e'_g$ bands, $W_{\rm ortho}\approx 2$~eV, 
is considerably smaller than in the tetragonal case.  
On the whole, therefore, the noncubic distortions are significantly smaller
for the tetragonal interlayer structure than for the orthorhombic bulk form
of LaTiO$_3$. This difference is also apparent in the shape of the density 
of states since the tetragonal distributions shown in Fig.~1 resemble more
closely the characteristic $t_{2g}$ density of cubic SrVO$_3$ (width 
$W_{\rm cubic}=2.8$~eV ~\cite{pavarini}).

\begin{figure}[t]
  \begin{center}
  \includegraphics[width=6.0cm,height=8cm,angle=-90]{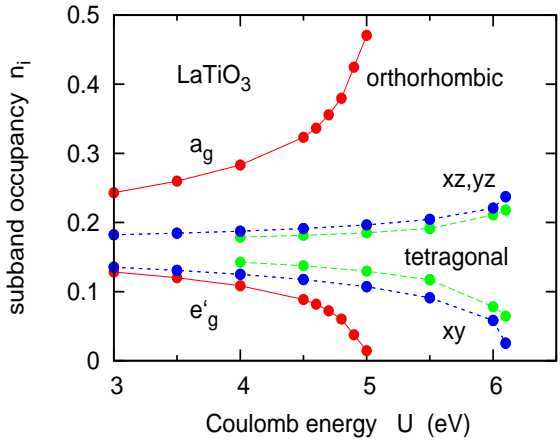}
  \end{center}
\vskip-4mm
\caption{(Color online) 
Ti $3d$ subband occupancies of LaTiO$_3$ as a function of local Coulomb 
energy. Solid (red) curves: orthorhombic bulk geometry, indicating
metal insulator transition at $U\approx5$~eV.
Short-dashed (blue) curves: tetragonal bulk geometry,
long-dashed (green) curves: tetragonal slab geometry,
both indicating Mott transitions at $U>6$~eV. 
Since the crystal field splitting has opposite sign in the orthorhombic
and tetragonal structures, the orbital polarization between singly
degenerate $a_g$ ($d_{xy}$) and doubly degenerate $e'_g$ ($d_{xz,yz}$) 
states is reversed.   
}\end{figure}

We now show that the combined influence of these differences: the wider 
band width, the weaker orbital polarization, and the opposite sign of this 
polarization, implies that LaTiO$_3$ in the tetragonal superlattice geometry
is a strongly correlated metal, rather than a Mott insulator as in its
orthorhombic bulk form.    

To evaluate the electronic properties of LaTiO$_3$ in the presence of strong 
local Coulomb interactions we use multi-band ED/DMFT~\cite{perroni}. Full
Hund exchange is included, with the exchange energy $J=0.65$~eV held fixed 
at the value appropriate for the bulk material~\cite{pavarini}. To locate 
the Mott transition, the intra-orbital Coulomb energy $U$ is varied, and 
the inter-orbital Coulomb interaction is given by $U'=U-2J$. The temperature 
is $T=20$~meV. 

\begin{figure}[t]
  \begin{center}
   \includegraphics[width=5.0cm,height=8cm,angle=-90]{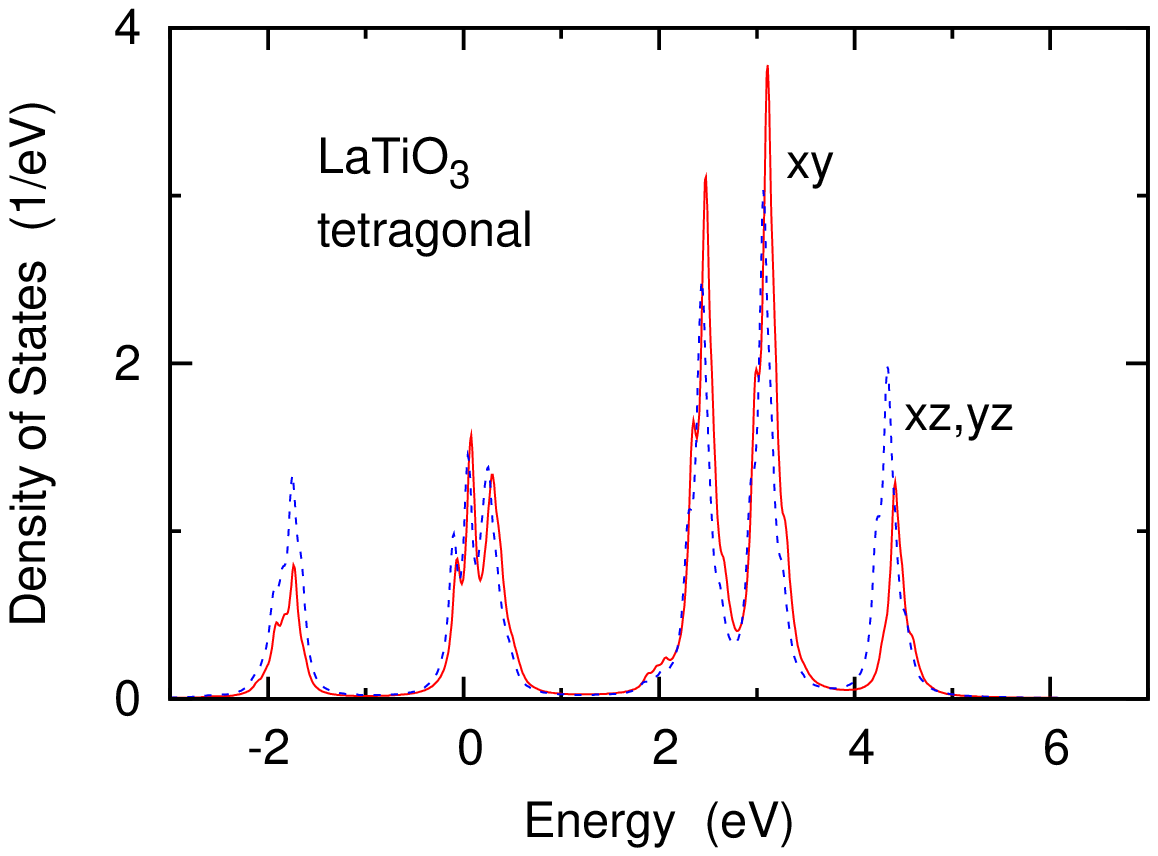}
   \includegraphics[width=5.0cm,height=8cm,angle=-90]{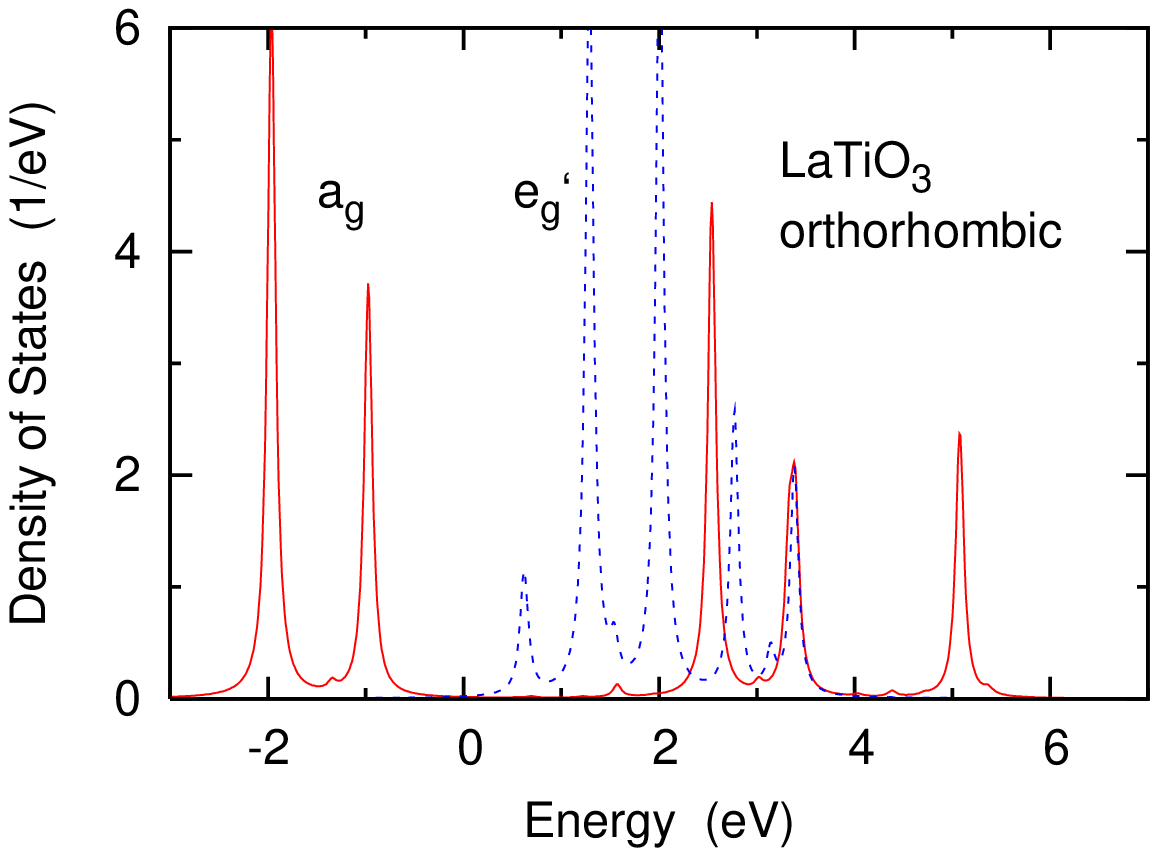}
  \end{center}
\vskip-4mm
\caption{(Color online) 
 LaTiO$_3$ $3d$ $t_{2g}$ spectral distributions. 
Upper panel: tetragonal superlattice geometry, $U=5.5$~eV,  
indicating strongly correlated metallic behavior. 
Lower panel: orthorhombic bulk geometry, $U=5$~eV, 
indicating Mott insulating behavior.  
Solid (red) curves: singly degenerate $d_{xy}$ ($a_g$) states, 
dashed (blue) curves: doubly degenerate $d_{xz,yz}$ ($e'_g$) states; $E_F=0$.
}\end{figure}

Fig.~3 shows the $a_g,\,e'_g$ subband occupancies as a function of Coulomb 
energy. For the orthorhombic geometry, nearly complete orbital polarization
is reached at about $U=5$~eV, which agrees with previous DMFT results using 
the Quantum Monte Carlo technique as impurity solver~\cite{pavarini}. 
The spectral distributions (see Fig.~4) show that the nearly half-filled 
$a_g$ band then consists of lower and upper Hubbard peaks whereas the $e'_g$ 
bands are pushed above the Fermi level~\cite{pavarini,doping}.      
In the case of the tetragonal structures, the situation is qualitatively 
different since the crystal field is weaker and has opposite sign. 
Now the enhancement
of the orbital polarization via the Coulomb interaction implies that the
$d_{xy}$ band is pushed above the Fermi level and that the 
degenerate $d_{xz,yz}$ bands are driven to $1/4$ occupancy. 
As a result of this reversal of crystal field, and because of its weaker
magnitude (i.e., weaker orbital polarization in the uncorrelated limit),
the metal insulator transition occurs above $U=6$~eV. Since the local   
Coulomb energy for Ti is approximately $5$~eV~\cite{pavarini}, 
this result implies that in the tetragonal geometry LaTiO$_3$ is on the 
metallic side of the Mott transition. As mentioned above, we have neglected
small contributions to the tetragonal density of states in the region of the 
empty $e_g$ bands. Inclusion of these states would increase
the tetragonal $t_{2g}$ band width and reinforce the metallic behavior.  

As can be seen in Fig.~3, the Mott transition for the tetragonal slab geometry
occurs at slightly larger $U$ than for the tetragonal bulk. This shift is a
consequence of the smaller orbital polarization in the former case, associated 
with the weaker deviations of the density of states components from cubic 
symmetry, as is evident also from the results shown Fig.~1.     
     
In Fig.~4 we compare the LaTiO$_3$ $t_{2g}$ spectral distributions for the 
tetragonal slab and orthorhombic bulk crystal structures. Since we are 
concerned here only with the difference between metallic and insulating behavior, 
it is sufficient to use the ED cluster spectra which can be evaluated at real 
$\omega$, without requiring extrapolation from imaginary Matsubara frequencies. 
In the orthorhombic case, the system is already insulating at $U=5$~eV, 
with an excitation gap formed between the lower Hubbard peak of the $a_g$ 
band and the empty $e'_g$ bands, in agreement with Ref.~\cite{pavarini}. 
The tetragonal slab geometry, on the other hand, is still metallic at
$U=5.5$~eV, with conduction states near $E_F$ from $d_{xy}$ 
and $d_{xz,yz}$ bands. 

In the DMFT calculations discussed above we have focused on the central 
Ti plane in the tetragonal LaTiO$_3$ layer. Ti planes at the interface 
with SrTiO$_3$ are also metallic because of the nominal $3d^{0.5}$ occupancy
~\cite{ohtomo,okamoto.04}. The mutual enhancement of both effects ensures 
that the entire LaTiO$_3$ layer is metallic. 

The above results demonstrate the extreme sensitivity of the electronic 
properties of perovskites to small changes in key parameters. In the
present case, the slightly modified crystal structure caused by the deposition
of LaTiO$_3$ onto SrTiO$_3$ gives rise to a subtle, but important change in
its single-electron spectral distribution, which, in turn, has a profound 
effect on the properties of the correlated electron system. For an adequate 
interpretation of experimental data it is therefore of great importance to 
have accurate structural information. On the theoretical side, it is crucial
to include the full details of the single-particle electronic properties
for a given lattice geometry and to account for the orbital degrees of 
freedom in the multi-band many-body calculation.

In summary, we have shown that in LaTiO$_3$/SrTiO$_3$ superlattices the 
change from orthorhombic to tetragonal crystal structure enforced by the
SrTiO$_3$ substrate causes a fundamental change in the electronic properties
of LaTiO$_3$. Instead of being a Mott insulator as in the bulk geometry,
it is now a highly correlated metal. Thus, the metalicity observed in these
heterostructures is not only associated with the interfaces but with the
entire LaTiO$_3$ layers. It would be very interesting to use transmission 
electron microscopy to determine the LaTiO$_3$ layer thickness at which the
orthorhombic structure begins to be energetically favorable. Mott insulating
behavior should then emerge in the central regions of these LaTiO$_3$ layers
at sufficiently large thickness.

\bigskip

\end{document}